\documentclass{aastex62}

\newcommand\aastex{AAS\TeX}

\makeatother

\submitjournal{ApJ}
\shorttitle{\aastex\ Local oscillations}
\shortauthors{Robert Sych et al.}

\begin{document}

\title{Properties of local oscillations in the lower sunspot atmosphere}

\correspondingauthor{Robert Sych}
\email{sych@iszf.irk.ru}

\author[0000-0002-0786-7307]{Robert Sych}
\affil{Institute of Solar-Terrestrial Physics SB RAS, Irkutsk 664033, Russia}
\author{Yuzef Zhugzhda}
\affiliation{Pushkov Institute of Terrestrial Magnetism, Ionosphere and Radio Wave Propagation RAS, Troitsk 108840, Russia}
\author{Xiaoli Yan}
\affiliation{Yunnan Observatories, Chinese Academy of Sciences, Kunming 650011, China}

\begin{abstract}
We present a study of wave processes in the sunspot region NOAA 12670 on 2017 August 10 
observed by the Goode Solar Telescope in the TiO 7057\AA ~and H$\alpha$ 6563\AA ~spectral lines. 
To study the distribution of power oscillations and their dynamics, we applied 
the pixelized wavelet filtering (PWF) technique. For the first time, the spatial structure 
of oscillation sources as a footpoints of fine magnetic tubes, anchored in the sunspot umbra 
was obtained. We found that at the chromosphere level, emission variation is a combination 
of numerous independent oscillations located in the sources with small angular size. 
Their spatial shape varies from dots and cellular in the umbra to filaments in the penumbra. 
Each narrow spectrum harmonic corresponds to its source, without global correlation among themselves. 
There is weak background as low-frequency oscillation sources are distributed on whole umbra.
At the photosphere level we found regions with co-phased broadband oscillations of the whole umbra. 
Their spectrum includes the $\sim$3-min harmonic, whose maximal value is localized in umbral dots (UDs), 
and the low frequency part near the $\sim$5-min period. It is shown that the oscillation sources are 
displaced at different heights with increasing angular size. We assume that the observed spatial 
distribution of wave sources indicates the existence a slow subphotosphere resonator with a vertical 
magnetic field in the umbra and wave cutoff frequency due to inclination of the magnetic field line 
in the penumbra.
\end{abstract}


\section{Introduction} \label{sec:intro}
The first observations of sunspot oscillations
\citep{1969SoPh....7..351B, 1969SoPh....7..366W} showed that they
are most pronounced in the core of CaII chromospheric lines as
so-called umbral flashes (UFs). Study of this phenomenon was carried
out in a number of papers
\citep{1981A&A...102..147K,1983SoPh...87....7T,2000Sci...288.1396S,
2001ApJ...552..871L, 2003A&A...403..277R, 2007PASJ...59S.631N,
2007A&A...463.1153T}. Umbral flashes are also associated with the
phenomenon of running waves in the penumbra, which is observed in
the H-alpha and He lines \citep{2007ApJ...671.1005B}, as well as
CaII \citep{2013A&A...556A.115D} as symmetric spatial structures
moving in the radial direction from the umbral center the outer
boundary of the penumbra \citep{2000A&A...355..375T,
2003A&A...403..277R}. Propagating waves are non-stationary, with a
change in oscillations power, both in time and in space
\citep{2010SoPh..266..349S}. This leads to a significant periodic
modulation of the propagating 3-min waves. A possible response of
such modulation is the appearance of both low-frequency wave trains
and separate brightening maxima emission as umbral flashes in the
footpoints of the magnetic tubes \citep{2018A&A...618A.123S}.

According to \cite{1977A&A....55..239B}, the low-frequency waves
generated at the sub-photospheric level (p-mode) propagate through
the natural waveguides as a concentration of magnetic elements -
sunspots and pores. Their oscillation period can be modified by wave
cutoff frequency. It is shown \citep{1977A&A....55..239B,
1984A&A...133..333Z} that oscillations with a frequency below the
magnetoacoustic cutoff frequency are quickly damped. The main factor
affecting the cutoff frequency is the inclination of the field lines
where the wave propagation is observed. Thereof, we observe 5-min
oscillations in diverging magnetic fields both in the chromosphere
(spicules, \citep{2004Natur.430..536D}) and in the corona (loops of
active regions, \citep{2005ApJ...624L..61D, 2009ApJ...702L.168D}).
The study of low-frequency oscillations in the upper levels of the
solar atmosphere \citep{2009ASPC..415...28W, 2009ApJ...697.1674M,
2011SoPh..272..101Y} confirmed the assumption that their appearance
at such heights is a consequence of channeling waves in inclined
magnetic fields. The observed propagation speed of the disturbances
indicates their nature as slow magnetoacoustic waves
\citep{2009A&A...505..791S, 2012SoPh..279..427K}.

It was assumed that oscillations in sunspots to originate due to the
penetration of five-minute oscillations from the convection zone.
After transformation into slow MHD waves, the oscillations propagate
upward into the photosphere, the chromosphere, and the corona.
Spectral analysis of observations showed the presence of five-minute
oscillations in the photosphere and three-minute oscillations in the
chromosphere and the corona. Many numerical experiments were carried
out within this approach \citep{2015LRSP...12....6K}.

At the photosphere level the 3-min sunspot oscillations are very low
and not visible due to negligible amplitudes compared to the
five-minute oscillations. For example, \cite{1987SoPh..112...37B}
could not detect their manifestations at the photosphere.
\cite{2007PASJ...59S.631N} showed that the oscillation power at all
frequencies significantly drops in the sunspot umbra. At the same
time, the first information about the possibility to register these
variations appeared. In  \cite{2008ESPM...12.2.38K}, the 3-min
oscillations at the photosphere level were detected for the first
time, and they were compared with the chromospheric data. It is
shown that the localization of the maximum power of oscillations at
the chromosphere level coincides in space with a minimum at the
photosphere level.  However, these observations with a spatial
resolution of $\sim$1\arcsec ~did not allow distinguishing the
two-dimensional structure of the oscillation sources. The 3-min
oscillations were interpreted in the framework of the chromospheric
resonator hypothesis. But observations \citep{1984MNRAS.207..731Z,
2001SoPh..202..281S} questioned this hypothesis by taking into
account the presence of local oscillations. The central problem was
the presence of powerful local three-minute oscillations detected in
the umbral dots at the photosphere level \citep{2012ApJ...757..160J,
2017AN....338..662E}.

In the chromosphere, oscillations become visible due to filtering
the five-minute oscillations and a sharp decrease in plasma density
compared to the photosphere. Local three-minute oscillations cannot
be explained in the framework of the hypothesis about penetration
and  transformation of the P-mode oscillations. This is because
P-mode effectively transformed only into large-scale oscillations
covering the whole sunspot, and do not transform into small-scale
local oscillations \citep{2014AstL...40..576Z}. This has been shown
within the monolithic sunspot model. Within the cluster sunspot
model comprising a set of separated magnetic tubes, it is impossible
to explain the local three-minute oscillations. The absorption of
P-mode in magnetic tubes is greatly weakened, which is considered as
a reason for the weakening disturbances in sunspots compared to the
surrounding solar atmosphere \citep{2014ApJ...796...72J}. This
absorption rapidly decreases with frequency, which excludes the
presence of powerful local three-minute oscillations at the
photosphere level.

\cite{1984A&A...133..333Z} and \cite{2014AstL...40..576Z} proposed
an alternative model of local oscillations, based on the possibility
of the existence of a sub-photospheric resonator for slow waves. An
excellent candidate for structures where   the resonance is possible
through convective jets (plumes) in the sunspots and was found in
the set of numerical experiments \citep{2006ApJ...641L..73S}. These
jets lead to the emergence of umbral dots, where the three-minute
oscillations were detected. The proposed model brings together such
key problems of the sunspot theory as the convective energy transfer
in sunspots; the presence of umbral bright dots; three-minute
oscillations; powerful brightening as umbral flashes; active
phenomena in the chromosphere and the sunspot corona under the UF
effect.

In this paper, we analyzed the fine spatial structure of wave
sources in the sunspot obtained at the  photosphere and chromosphere
level, and examined their relationship with active energy phenomena
like umbra flashes and umbra dots. For spatial localization of the
sources, the pixelized  wavelet filtering technique was used.
The article is structured as follows: in Section 1 we give an
introduction on the topic of work and present the objects of study;
in Section 2, we describe observation and data processing; in
Section 3 the data analysis with   results and discussion on the
physical processes connected  with the considered wave phenomena at
different heights is presented, and Section 4 contains the
conclusions of the obtained results.

\section{Observations}

The two optical channels centered at the H$\alpha$ line core (6563\AA)
and TiO (7057\AA) of  the Goode Solar Telescope (GST)
\citep{2011ASPC..437..345C}, located at the Big Bear Solar
Observatory (BBSO), were used to observe the chromospheric and
photospheric images of the active sunspot region NOAA 12670 (S06
W47) on August 10, 2017 (18:18-19:11 UT). The observations are
carried out using a high-order adaptive optical system with 308
sub-apertures. We used the Broadband Filter Imager (BFI) for
obtaining the TiO data and the Visible Imaging Spectrometer (VIS)
for obtaining the H$\alpha$ data. The field of view (FOV) of both
instruments is 70\arcsec. The pixel scale is 0.0295\arcsec ~for the
H-alpha center and 0.0342\arcsec ~for the TiO images with spatial 
resolution 0.1\arcsec ~and 0.11\arcsec ~respectively.  The cadence
for the H$\alpha$ images is 18 s and 13 s for the TiO channel. We observed
seven wavelength points $\pm$0.0, $\pm$0.4, $\pm$0.8, and $\pm$1 of
the H-alpha line. The exposure times of H-alpha images were 7 ms, 9
ms, 15 ms, and 20 ms respectively. All data are corrected using
standard software provided by the BBSO.

\begin{figure}
\begin{center}
\includegraphics[width=13.0 cm]{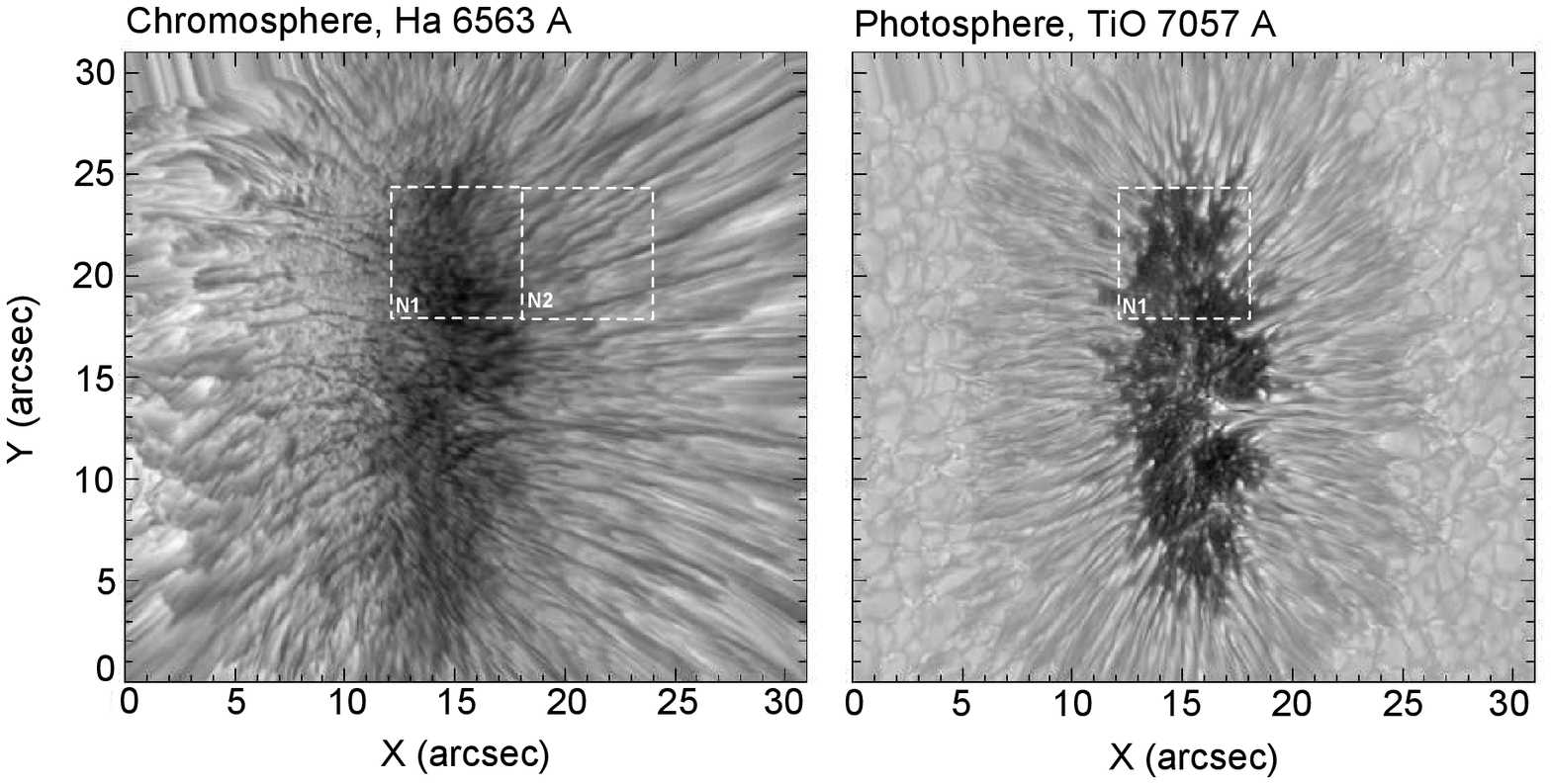}
\end{center}
\caption{Image of the NOAA 12670 sunspot active group on 2017 August 10 at 18:18 UT, obtained with the GST at BBSO in the H$\alpha$ 
6563 \AA ~line (left panel) at the chromosphere level, and in the TiO 7057 \AA ~line (right panel) at the photosphere level. White numbered rectangles indicate the studied areas, N1 and N2. Brightness is on a log scale. Spatial coordinates are in arcsec.}
\label{1}
\end{figure}

We applied the PWF technique \citep{2008SoPh..248..395S} for spectral data preparation. This method was previously widely used by authors in studies of sunspot oscillations \citep{2010SoPh..266..349S, 2014A&A...569A..72S, 2015A&A...577A..43S, 2018A&A...618A.123S}. 
The numerical method is a generalisation of the wavelet transform of 3D datacubes. Temporal signal of each spatial pixel is wavelet transformed with Morlet mother function (\cite{1998BAMS...79...61T}), which results in the power, amplitude, and phase 4D data cubes (two spatial dimensions, time, and the frequency). The obtained data can be processed according to a specific request. For example, selecting a certain spectral component or integrating over a certain narrow spectral range makes a 3D narrowband datacube that consists of a sequence of narrowband maps. 

We reconstructed the time signals using a spectral filter as a running window with the narrow band from $P_{i}/1.25$ to  $P_{i}*1.25$, and step 0.1 min, where $P_{i}$ is the current value of the period. The range of periods was from 1 min to 8 min. For each reconstructed signal, the values of amplitude variations were calculated. Repeating a similar procedure for all image points, we prepared a series  surfaces as narrowband images of oscillation sources. To show the global 1D spatial-periods dependence along the selected spatial axis, we used scanning of the narrowband sunspot images with a coordinate-period diagram preparation. A similar technique was applied in \cite{2014A&A...569A..72S} to study oscillations using SDO/AIA data in the UV range. 

To conveniently present radially propagating wave fronts, and to calculate their spectral characteristics, all sunspot images were rotated counter-clockwise by 12$^{\circ}$ for vertical positioning. The space-averaged coordinate-period diagrams were obtained, which allowed obtaining a 1D distribution of the oscillation sources. To localize their positions in space, we also obtained the 2D structure as an images with calculated oscillation periods and power.

Based on the PWF technique, we developed a method of period and power color maps preparation. For each spatial point of a narrowband image cube, we constructed the period profiles and calculate their spectral peak and corresponding period oscillations. We assumed that these values should be no less than half of the maximum spectral power in a given period band. The points with the maximum oscillations power were selected using a certain color in accordance with the color table. This table defines the specified sequence of color depending on the periods. The points with a spectral peak below the specified level were chosen as zero, with no oscillations, and not displayed on the color map. The obtained details as repeating points of the same color form the oscillation sources with the same periods. A similar mapping method was previously developed in \citep{2007SoPh..241..397N} by using Fourier transforms for spectral decomposition of signals.

\section{Data analysis and discussion}

Figure ~\ref{1} shows images of the sunspot active group NOAA 12670 obtained on August 10, 2017. The start observations is at 18:18 UT, and duration is 53 min. The spatial details in the H$\alpha$ (left panel, chromosphere) and the TiO  (right panel, photosphere) shows the pronounced processes of emission variation as running waves, umbral flashes and bright dots. Broken squares indicate the study areas N1 (umbra) and N2 (penumbra) with an angular size of 6x6\arcsec.  The number of images obtained in the TiO line was 247 frames and 178 frames in the H$\alpha$ spectral line. The duration and cadences of observation allowed identifying oscillations with periods ranging from $\sim$1 min to $\sim$20 min.

\begin{figure}
\begin{center}
\includegraphics[width=15.0 cm]{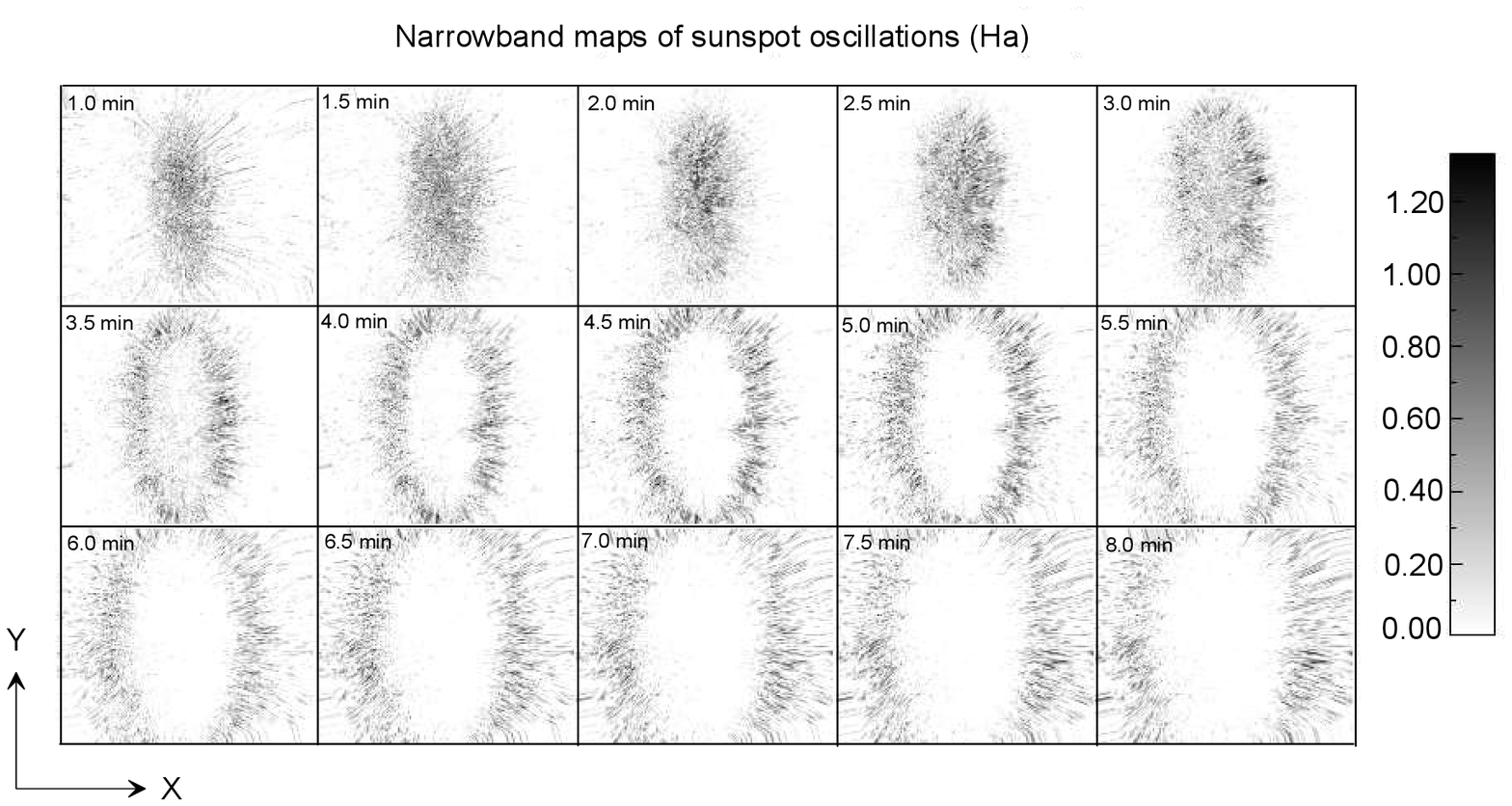}
\end{center}
\caption{Sunspot narrowband maps of oscillation power for the H$\alpha$ 6563 \AA wavelength. Period values are in minutes.}
\label{2}
\end{figure}

\subsection{1D fine structure of sunspot oscillations}

Filtering the data cubes through a running window with average intensity subtraction, we obtained a set of sunspot narrowband images as ellipses that, as the period grows, transform into expanding rings. We can see (Fig. ~\ref{2}) that, for high-frequency oscillations at chromosphere level (H$\alpha$), sources with a period of less than 3 min are localized in the sunspot umbra, and the area that they occupy decreases with  period reduction. This regularity agrees well with the earlier studies of sunspot oscillations \citep{2012ApJ...756...35R, 2012ApJ...757..160J, 2014A&A...561A..19Y}. We can see also a fine structure as dotted and extended small-size patches that fill the umbral central part.

\begin{figure*}
\begin{center}
\includegraphics[width=15.0 cm]{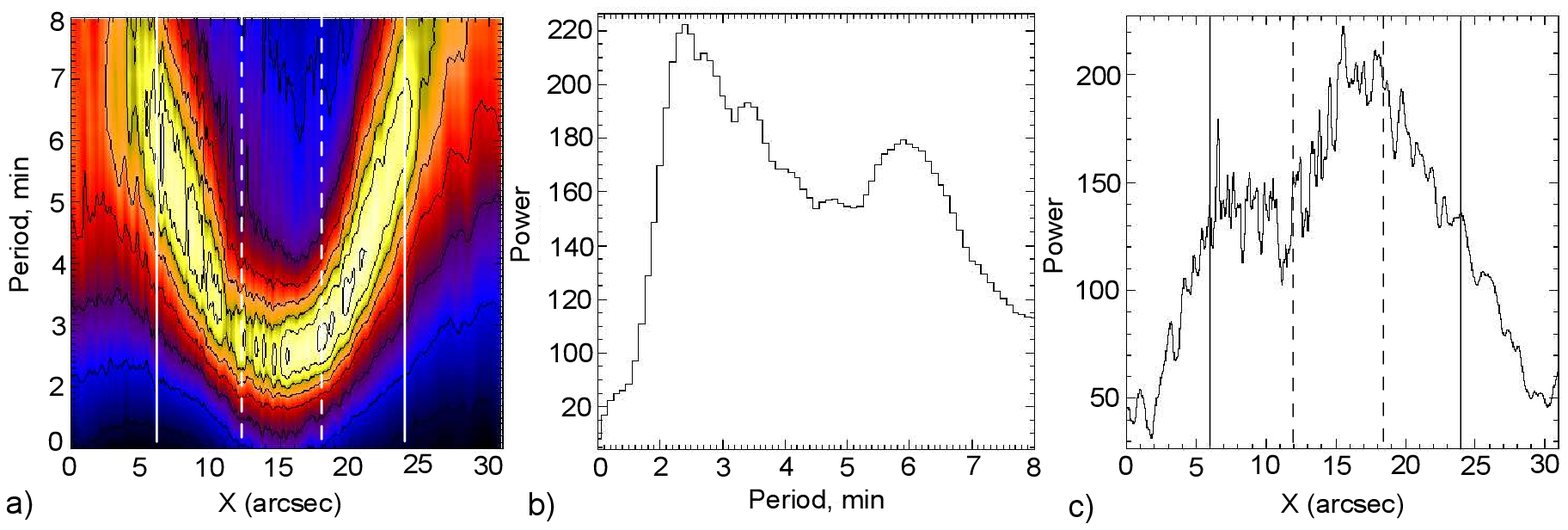}
\end{center}
\caption{Spectral distribution of the oscillations power for the sunspot in the H$\alpha$ (6563 \AA, chromosphere). (a) Coordinate-period oscillation diagram. (b) Oscillation spectrum. (c) 1D spatial distribution of the sunspot oscillation power. Dashed and continuous lines show the umbral and penumbral boundaries, respectively. Oscillation periods are in minutes, and spatial coordinates in arcsec.}
\label{3}
\end{figure*}

\begin{figure*}
\begin{center}
\includegraphics[width=15.0 cm]{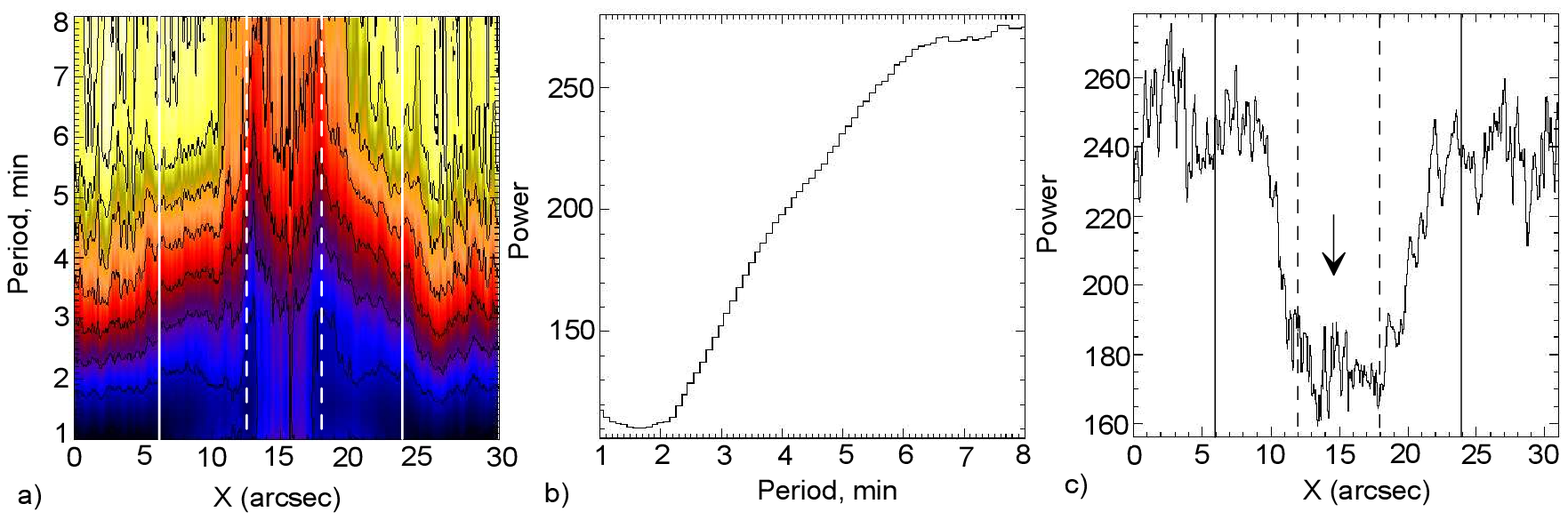}
\end{center}
\caption{Same as Fig.~\ref{3} in the TiO (7057 \AA, photosphere). The arrow shows the fine umbral oscillations increase.}
\label{4}
\end{figure*}

Using a set of narrowband images (Fig. ~\ref{2}) we can obtain a spectral distribution of the averaged oscillations power in the sunspot (Fig. ~\ref{3},~\ref{4}). For this we scanned the images along X-axis for each point of the Y-axis. Then, the resulting set of scans was averaged. The average was performed due to spatial inhomogeneities of the oscillations. During our analysis we have excluded the upper and bottom boundaries of the images from the averages. The obtained 1D scans for each period composed a coordinate-period diagram (Fig.~\ref{3}a) as a two diverging ellipse-like spatial details that trace the sunspot oscillation power. The angular size of these details increases with period and displaces the penumbral boundaries. This indicates the existence of diverging magnetic field lines, along which waves propagate upward, into the corona \citep{2012A&A...539A..23S}.

We can see (Fig.~\ref{3}a) that the high-frequency part of the sunspot oscillations is located within the umbral boundaries with $\sim$5\arcsec ~size, marked by vertical dash lines. Calculation the periods where the power is maximal gave us the oscillations spectrum   (Fig.~\ref{3}b). The same procedure but for coordinates gave us the spatial distribution of power oscillations (Fig.~\ref{3}c). There are two spectral peaks: one near the 3-min period, and the other within 5-7 minutes. For convenience, we will term the oscillations in the 2-4 min high-frequency range as the $\sim$3-min oscillations, and in the 4-8-min low-frequency range as the $\sim$5-min oscillations. The  maximal power in the umbra is related to the 3-min periodicity.

In Fig.~\ref{3}a we can see the sources with small $\sim$1-2\arcsec ~angular size. This indicates that, even at the strong averaging over the whole sunspot, a fine spatial structure is detected in the umbra with minimal angular size. In quiet regions, the low-frequency component dominates.

Using the data processing similarly to the chromosphere level we obtain the photosphere power distribution of oscillations. Figure ~\ref{4} shows the obtained coordinate-period diagram (a), oscillation spectrum (b), and spatial distribution of the spectral power (c).

Comparing the obtained coordinate-period diagrams at different heights we obtained a strong difference from each other. The chromospheric maximum in the sunspot umbra (Fig.~\ref{3}c) coincides with minimum oscillations at the photosphere level (Fig.~\ref{4}c). At the photosphere we do not observe growth of oscillation period from the umbra center. We see (Fig.~\ref{4}a)  that the oscillations have a broadband character. The main power is concentrated in a low-frequency range (Fig.~\ref{4}b), which is related to the global $\sim$5-min oscillations as a p-mode.

As we move toward the umbral center, we observe a smooth lowering of the background oscillations. This trend is clearly visible in the penumbra and has a symmetric character. Then, within the umbral boundaries, the oscillation power starts to grow in a broad range (Fig.~\ref{4}a) but for fine structures. This dependence seen clearly on the 1D profile of the spectral power (Fig.~\ref{4}c), where the arrow marks the local increases.

We found a fine spatial structure of sources in the umbra, as well as at the chromosphere level. The oscillations do not occupy a narrow range of periods like 3-min sources at the chromosphere (Fig.~\ref{3}a), but a broad range extending up to 8 minutes. The shape of the sources shows a small size and stable localization. The averaged spectrum (Fig.~\ref{4}b) does not have significant oscillation peaks. There is only a growth of oscillation power with period. The umbral oscillations are very weak and not detected in the full signal.

\subsection{Spatial distribution and dynamics of oscillations}

Recently the theory of 3-min oscillations significantly changed because of the results obtained with high-resolution. The oscillations are a result of the penetration of the p-mode from the surrounding quiet atmosphere into a sunspot \citep{2015LRSP...12....6K} as a tail of a broadband oscillations spectrum in a quiet atmosphere. Due to the cutoff of the spectrum, only a three-minute periodicity remains in the temperature minimum of the chromosphere. In the photosphere, weak 3-min oscillations are not visible against powerful 5-min oscillations. However, the 3-min oscillations localized in small areas appeared to be visible as a spatial resolution of the observational instrument was increased \citep{2012ApJ...757..160J, 2014AstL...40..576Z, 2015ApJ...812L..15K, 2017ApJ...836...18C, 2018RAA....18..105Z}. The obtained umbral oscillations show the different spatial and spectral localization depending on height. Studying the GST BBSO high-resolution  observational data from different heights of the sunspot atmosphere enabled us for the first time to understand this fine structure.

\subsubsection{Chromosphere level}

The detected structure of the oscillation sources (Fig.~\ref{3}a) was obtained by averaging power over the whole sunspot. This is   not sufficient to obtain the oscillations' fine structure. To study the spatial location of the detected sources, we selected the umbral area N1 (Fig.~\ref{1}) with size 6x6\arcsec ~and obtained the 1D coordinate-period diagram. To study the 2D sources, we applied the PWF technique and obtained a set of narrowband images within the 1-8 min period range.

\begin{figure}
\begin{center}
\includegraphics[width=14.0 cm]{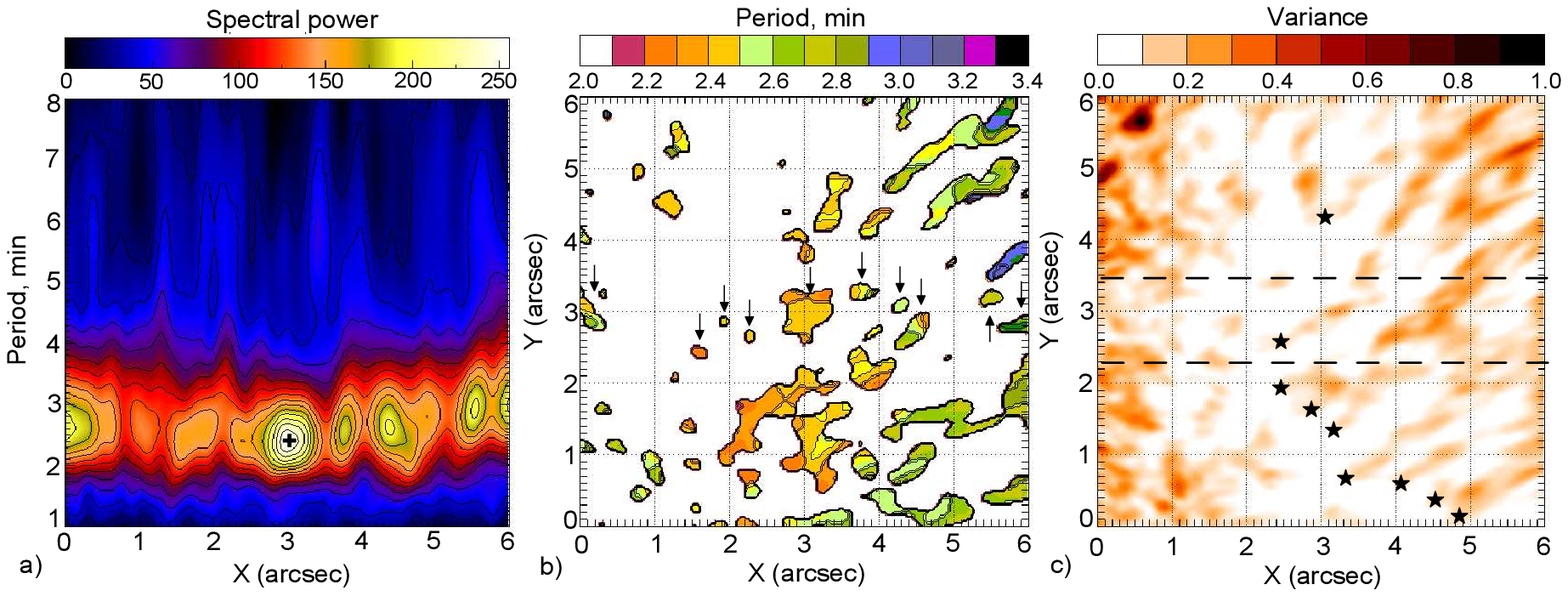}
\end{center}
\caption{(a) 1D spectral distribution of the oscillation power (coordinate-period diagram) of area N1 in the H$\alpha$ spectral line. (b) 2D color map of the oscillation periods. (c) Variance map of the intensity oscillations. Horizontal dash lines show the scanning region. Arrows indicate the local oscillation sources that correspond to powerful sources on the 1D diagram. The stars show the  beginning of the wave paths.}
\label{5}
\end{figure}

Figure~\ref{5} shows the obtained coordinate-period diagram, the 2D color map for the oscillation periods, and variance map of intensity oscillations. The horizontal dashed lines show the region of spatial scanning. The maximal periodicity is observed within the 2-4-min period range, with the maximum near 3 min (Fig.~\ref{5}a). Beyond this one, the oscillation power decreases. There is a fine ($\sim$0.2-0.5\arcsec) spatial structure of local sources with some inclination of the spectral power. A similar dependence, but more expressed, is observed for the whole sunspot (Fig.~\ref{3}a). We assume that the detected fine spatial structure is related to the sources of the $\sim$3-min oscillation harmonics.

To detect the spatial structure of the 3-min oscillation sources, we built a color map of the area N1 (Fig.~\ref{5}b). The level of the studied sources is limited to 50$\%$ of the power peak. We found that there is an expressed oscillation locality for all periods. The periodicity within 2-4 min represents a big number of independent oscillation sources with close amplitudes and periods. The arrows (Fig.~\ref{5}b) indicate bright sources within the scanning band on the coordinate-period diagram. We  observed only local oscillations in the identified sources. The size of the sources with $\sim$3-min periodicity varies from 0.2 to 0.7\arcsec. They have a high quality factor of oscillations and a weak coupling among themselves. The wave propagation occurs radially, along the selected directions.

\begin{figure}
\begin{center}
\includegraphics[width=9 cm]{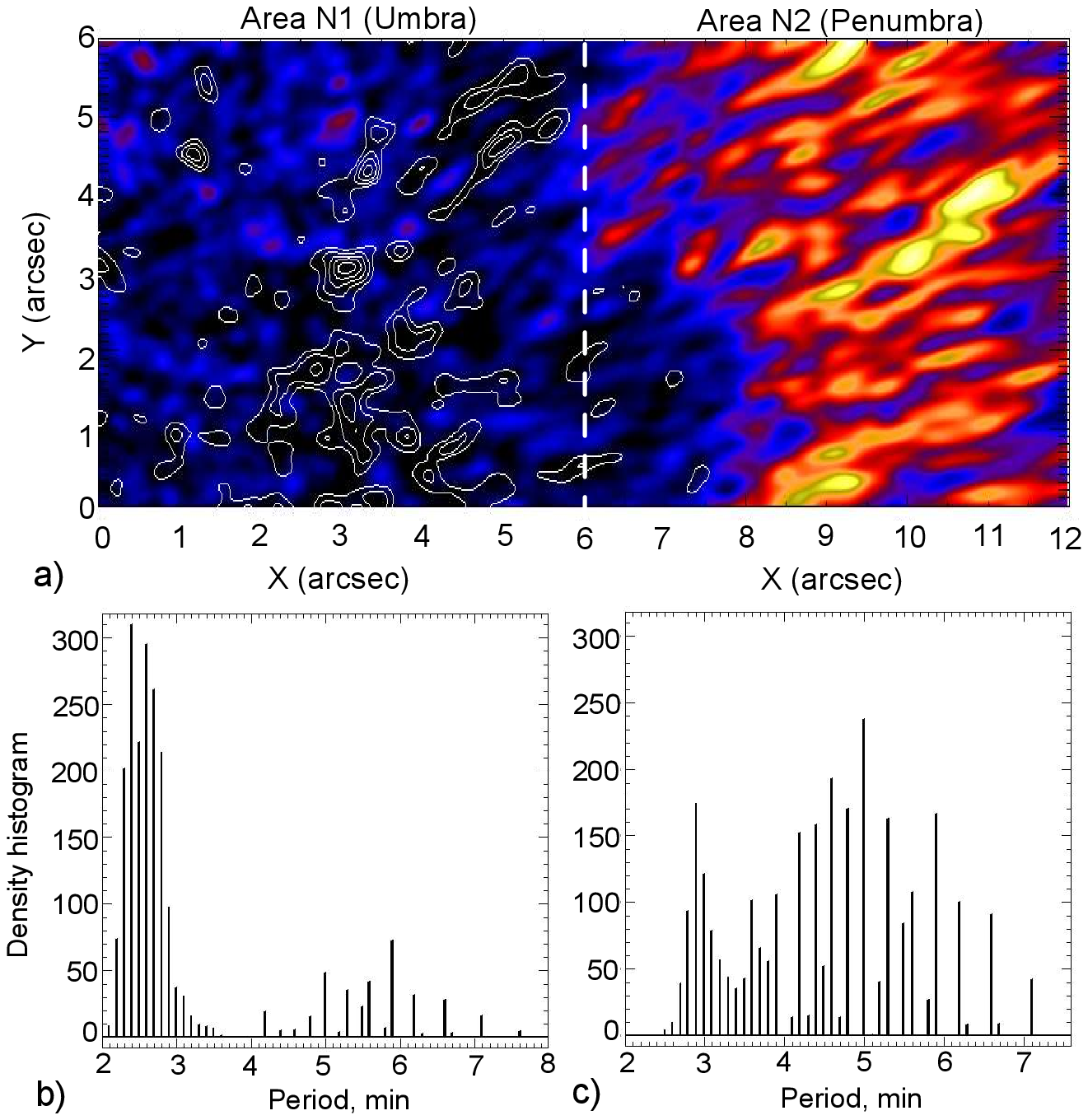}
\end{center}
\caption{(a) Spatial distribution of power oscillations for the highlighted areas N1 and N2. The  background is $\sim$5-min periodicity, and the contours are the $\sim$3-min oscillations. Histograms for the oscillation period spectral density are shown for areas N1 (b) and N2 (c). Period values are in minutes, and the size in arcsec.}
\label{6}
\end{figure}

The shape of sources changes from a cellular structure in the umbral center to filament one at its boundary, as the oscillation period grows. There are both dot sources with a very narrow spectral bandwidth, and extended structures, like cells and filaments. Some comprise a set of narrowband sources. The direction of the  the extended structures often agrees well with localization  of the cellular structures.

Since the intensity of each point in the temporal data cube varies greatly both in time and space during wave propagation, we prepared a variance map (Fig.~\ref{5}c) of intensity oscillations. We found a set of elongated filamentous structures (waveguides) along which we observed the waves propagation. The beginning of these structures coincides with the detected sources of $\sim$3 min oscillations and periodically observed bright details on the intensity map. This indicates that these spatial details are interconnected and possibly are the footpoints (cells or dots) of thin magnetic tubes	and waveguides (filaments) of propagating waves. 

It should be notes that the sources with small angular size, comparable with the angular resolution of the GST BBSO, are sources of periodic oscillations certain frequency and high power. This means that the high-frequency noise components, aperiodic in nature, will not be visible on the filtered images. All sources are well localized in space and time, and it power level is significantly higher than the noise level. 

The locality of oscillation sources demonstrates behavior not only the $\sim$3-min periodicity, but also for the low-frequency harmonics. To study the weak umbral $\sim$5-min spectral component, we prepared a color map for the full range of variation.  We chose area N2 to investigate the periodicity in the penumbra. The angular size is the same as for N1. Figure~\ref{6}a shows narrowband images of the area N1 (umbra) and the area N2 (penumbra) with power distribution within the $\sim$5-min period range. We overlap the $\sim$3-min oscillation spectral power as contours. The histograms for the oscillation spectral density distribution for each areas  were obtained (Fig.~\ref{6}b,c).

We found that in the 3-min range, the maximum oscillation power correlates well with the center of local cells and dot sources on the period color maps (Fig.~\ref{5}b). For extended structures, there are a few power peaks located along the sources. There are sources localized in the 5-8 min range,  like in the 2-4 min range (Fig.~\ref{6}a, background). The power histogram in the umbra (N1 area) shows a set of narrow harmonics within the 2-4 min periods with a significant peak near 2.5 minutes. The power of harmonics varies, depending of the period. Each harmonic corresponds to one and/or several narrowband sources presented on the period color map (Fig.~\ref{5}b). There is also a weak low-frequency spectrum part with the maximum near 5-6 minutes (Fig.~\ref{6}b).  For area N2, the main oscillations occur within the $\sim$5-min range. There is also a 3-min peak caused by overlapping the umbral region. The observed sources have mainly a filament shape and are placed radially from the umbral center.

Comparing the same spectral peaks shows different values in the
umbra and in the penumbra (Fig.~\ref{6}b,c). At the maximal value of
the $\sim$3-min component in the umbra, its penumbral value is
minimal. But, on the contrary, at the minimal power of 5-min
oscillations in the umbra, they are maximal in the penumbra. The
low-frequency sources are arranged among the high-frequency ones,
with no overlap. In the umbra, the shape of the sources with the
$\sim$5-min periodicity has a cellular structure which smoothly
transforms into a filament one in the penumbra.

\begin{figure}
\begin{center}
\includegraphics[width=10. cm]{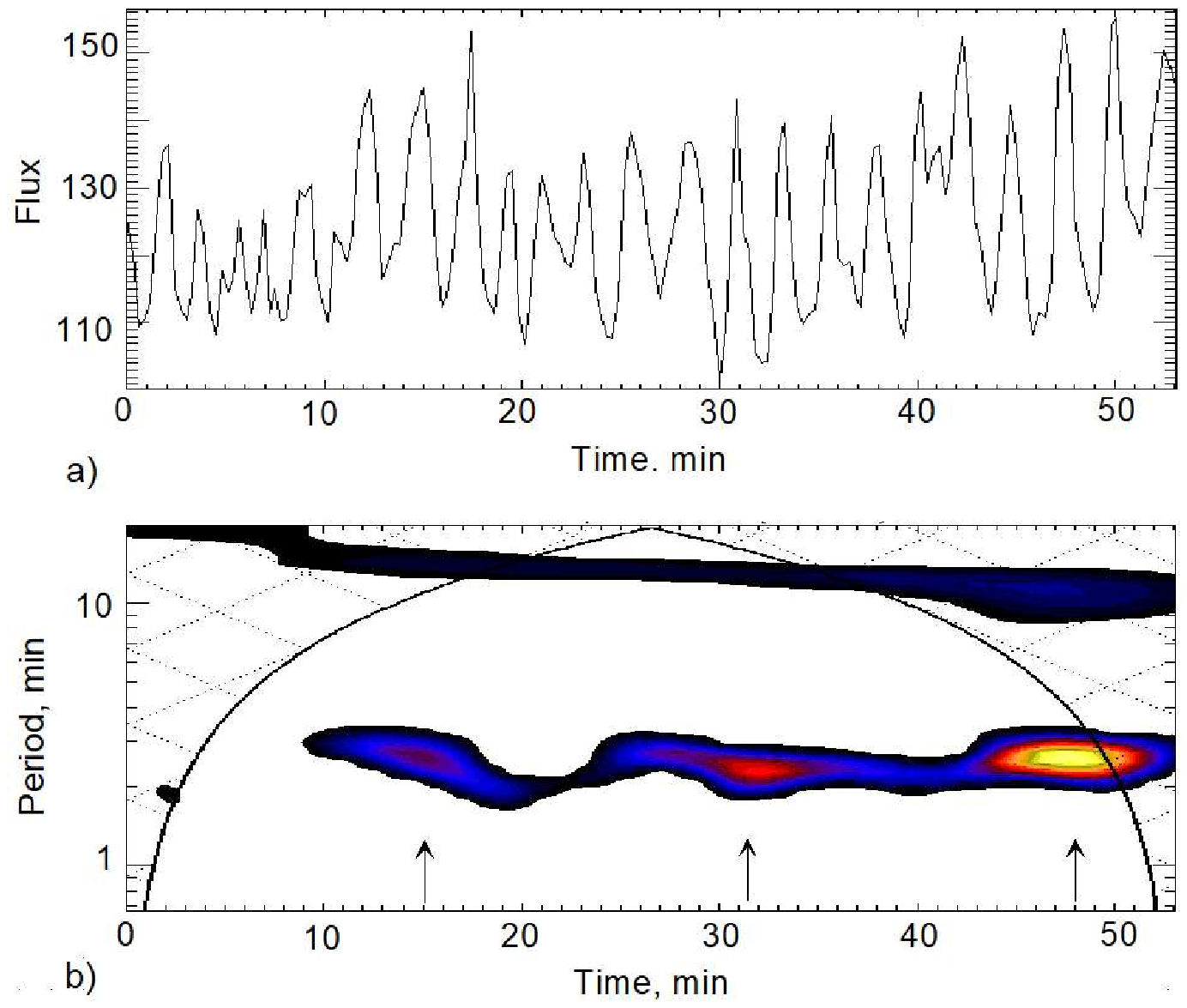}
\end{center}
\caption{(a) Time profile of the flux variation of detected
oscillation source as a cell in the center of area N1. (b)
Corresponding power wavelet spectrum. Arrows show the peaks of 3-min
oscillation trains.} \label{7}
\end{figure}

To study the temporal parameters of the detected oscillations, we
selected an extended source with the maximal amplitude. The cross
indicates the source on the coordinate-period diagram
(Fig.~\ref{5}a). This source coincided spatially with an oscillating
cell having an angular size of $\sim$0.5\arcsec ~and which is located
in the center of the color period map (Fig.~\ref{5}b).
Figure~\ref{7} shows the flux temporal profile obtained by
integrating the studied area, and its power wavelet spectrum.

We found that non-monotonous periodic oscillations prevail in the
source. To study them, we built a power wavelet spectrum with 90$\%$
significance. Figure ~\ref{7}b shows that the oscillations are
concentrated within a 2-4-min period range, and are modulated by
three low-frequency trains. The maxima of the revealed oscillation
period are $\sim$2.5 min (main oscillations) and $\sim$13 min (wave
trains).

A similar spectral analysis for the dot sources showed the presence
of oscillation trains, but without period drifts. There is only one
oscillation frequency within the envelope curve. We can assume that
the drifts in cells are caused by dynamics of a collection of the
various-period dots, where the periods vary with time during
increase of train power. For the cell we see (Fig.~\ref{5}b) the
several fine sources at the cell boundary, whose periodicity
variation may lead to the observed drifts. For dot sources, there is
only one modulation period without drifts.

The weak oscillations within $\sim$5-min periodicity are unstable
and vary with time. During the observational period, the $\sim$3-min
periodicity may turn into the $\sim$5-min one. The duration of the
components in the signal determines what period and power we will
observe in the spectrum. For powerful     $\sim$3-min oscillations
in the umbral cells and for $\sim$5-min oscillations in the
penumbral filament sources, variations that are stable in  time and
period amplitude are observed.

Figure \ref{2} and the coordinate-period diagram (Fig.~\ref{3}a)
show a symmetric increase of the oscillation period with distance
from the umbral center at the chromosphere level. High-frequency
oscillations are concentrated in the umbra, whereas the
lower-frequency component as expanding oscillating rings are located
in the penumbra (Fig.~\ref{2}b). A similar relation was described
earlier \citep{2008SoPh..248..395S, 2009ARep...53..957K,
2012ApJ...746..119R,  2013ApJ...779..168J, 2014A&A...561A..19Y}. This dependence can by
interpreted as variation of magnetic field inclination, and,
correspondingly, changes in cutoff frequency
\cite{1977A&A....55..239B}. The maximal power value with $\sim$3-min
periodicity (Fig.~\ref{3}c) is localized within the umbral
boundaries. This dependence is opposite to the umbral depression at
the photosphere level (Fig.~\ref{4}c).

The 3-min periodicity consists of the spatially separated sources
having a small angular size. There is a weak relationship among
themselves with an absence of in-phase oscillations. The size of the
sources is larger than in the photosphere. The waves propagate
radially from the center toward the penumbra along filament sources (Fig.~\ref{5}c), 
whose footpoints look like cells. On the color map the period decreases with distance 
to the umbral boundary (Fig.~\ref{5}b). A similar spatial-frequency dependence 
was found in \cite{2006A&A...456..689T, 2007A&A...463.1153T}. 
\cite{2014AstL...40..576Z} showed that 3-min umbra oscillations are local 
with the spectrum comprising tens of spectral lines. The oscillations are 
connected with small sunspot areas. The differences in the spectra of adjacent 
areas are insignificant.

The low-frequency wave trains (Fig.~\ref{7}) in sufficiently large
cells may be interpreted as increasing of wave activity in small
umbral areas. \cite{2012A&A...539A..23S} show that the period drifts
in the UV range coincide with the appearance of new fine structures
in the umbra with maximal power of 3-min oscillations. The
oscillation period varies with time. We assume that the period
drifts can explained  by spatial splitting of propagating waves
along the detected fine magnetic tubes with different physical
(temperature, density, field strength) and observational
(inclination, period) parameters  in the framework of the Parker
model. The simultaneous wave propagation along magnetic tubes with
different cutoff frequencies may explain the observed period drifts.

Oscillations in sunspot are the slow waves propagating along
magnetic field lines. P-modes during penetration into a sunspot
transform in fast waves with conversion into slow waves only at the
photosphere level. This process was explained  within a vertical
field and isothermal atmosphere model \citep{2014AstL...40..576Z,
2018RAA....18..105Z}. Only large-scale disturbances, comparable with
the umbra size, can originate as a result of p-mode conversion into
slow waves. This is a wave front which propagates toward the umbral
boundary while transforming into running penumbral waves (RPW).

\cite{2014AstL...40..576Z} show that 3-min oscillations in small UDs
originate due to exciting oscillations in the subphotospheric
resonator for slow waves. The existence of this resonator, predicted
by \cite{1984MNRAS.207..731Z}, was proved using a  thin-tube Roberts
approximation \citep{2006RSPTA.364..447R}. Within this model, a
mechanism of umbral flash origination was proposed
\citep{2018AstL...44..331Z, 2019AstL...45..177Z}.

We found that the traces from propagating wave fronts, which are
spirals or circles  \citep{2014A&A...569A..72S,
2016ApJ...817..117S}, appear not continuously, but consist of many
small details that brighten during the transition through the
sunspot as umbral flashes. The other observed detail is the
existence in the umbra of both 3-and 5-min oscillations
simultaneously. We assume that both phenomena are a consequence of
sunspot inhomogeneity. During propagation, the wave fronts interact
with these inhomogeneities. Umbral dots are also inhomogeneities. We
can suppose that they are related to the local variations in the
magnetic field geometry and accordingly the changes in local cutoff
frequency. Such complicated spatial-spectral structure of wave
fronts \cite{2012A&A...539A..23S} was shown early.

\subsubsection{Photosphere level}

The coordinate-period diagram (Fig.~\ref{4}a) showed the increase of
oscillation power in the fine umbra structures at the photosphere
level. We studied their distribution and parameters in area N1.
Similar investigations were performed for the same area at the
chromosphere level, and were described above. This allowed comparing
the observational parameters of the identified sources at various
heights of the sunspot atmosphere and their relationship.

\begin{figure}
\begin{center}
\includegraphics[width=15.0 cm]{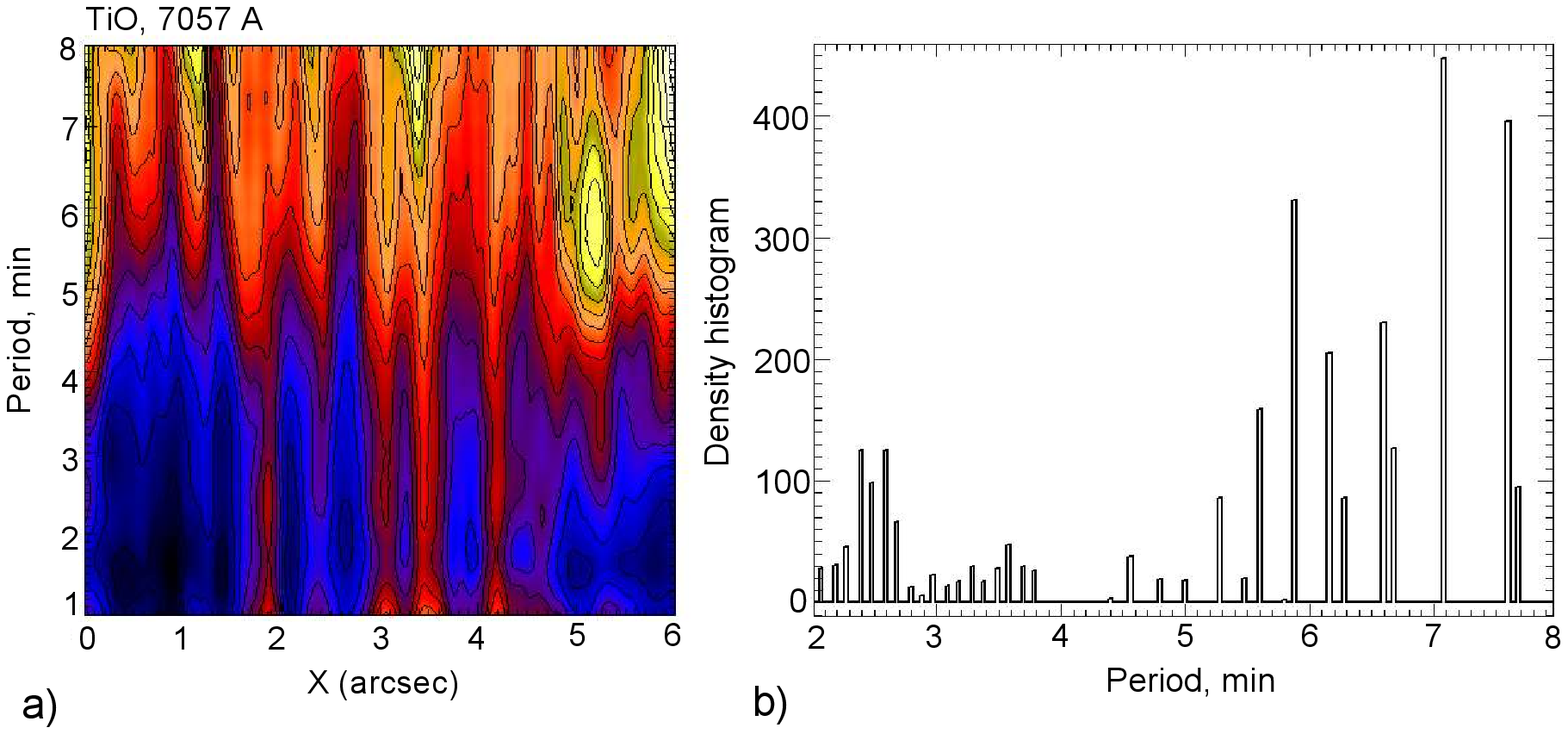}
\end{center}
\caption{(a) 1D spectral distribution of the power oscillations
(coordinate-period diagram) of area N1 in the TiO (7057 \AA). (b)
Histogram for the oscillation period spectral density. Period values
are in minutes, and size in arcsec.} \label{8}
\end{figure}

We calculated the coordinate-period diagram (Fig.~\ref{8}a) for the
central part of the N1 area. Similar data preparation for the H$\alpha$
spectral line (Fig.~\ref{5}a) was done previously. The obtained
diagram  shows the broadband sources as the extended spatial threads
filling the period range from 1 min up to 8 min. The angular size of
$\sim$0.2-0.3\arcsec ~for the identified sources is less than at
chromosphere level.

The histogram of spectral density shows (Fig.~\ref{8}b) that, like
at the chromosphere level (Fig.~\ref{6}b), there are two spectral
ranges, where the oscillations are present. The first range is near
the $\sim$3-min weak periodicity, the other one is near the
$\sim$5-min with the maximal oscillation power.

The maximum emission variation at the photosphere level is observed
outside the umbra and related to non-periodic plasma outflow in the
sunspot. In the umbra we see the periodic oscillations. These
oscillations in various parts of the N1 area showed simultaneous
temporal changes as standing waves. There is no running wave inside
the oscillating areas like at the chromosphere level.

\begin{figure}
\begin{center}
\includegraphics[width=8.5 cm]{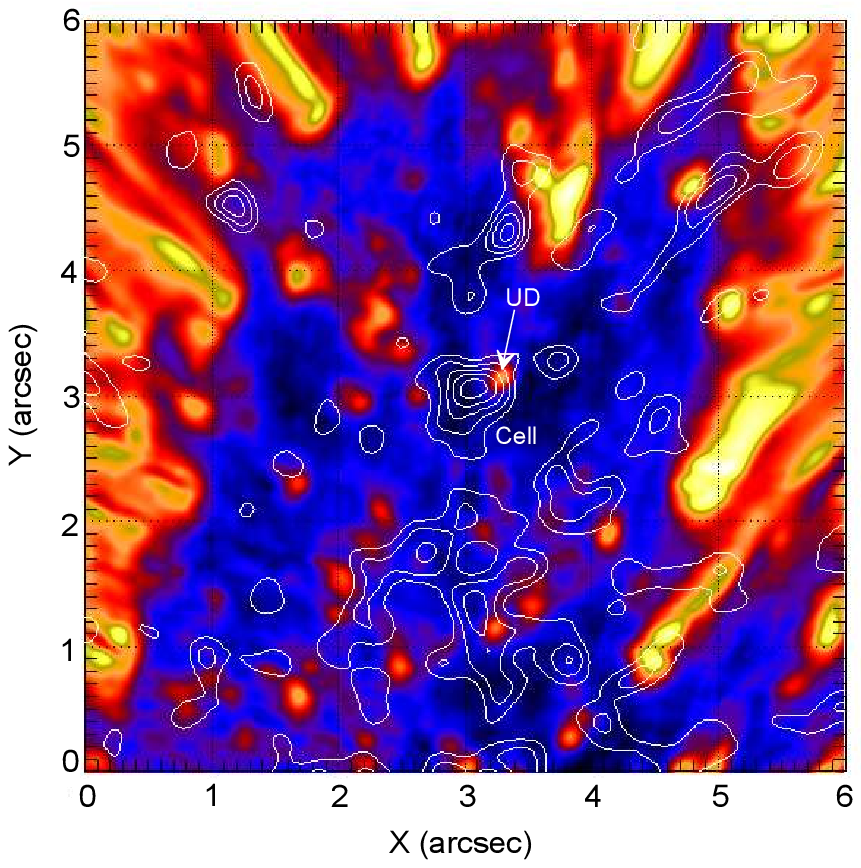}
\end{center}
\caption{Image of the umbra obtained in the TiO at 18:36 UT.
Contours show the location of $\sim$3 min oscillation sources in the
H$\alpha$ at the chromosphere level. Brightness is shown in a logarithmic
scale. An arrow indicates the studied UD.} \label{9}
\end{figure}

In the umbra we found the separated dots and extended sources as
brightenings of some parts of umbra (Fig.~\ref{9}, background).
These bright sources are related to the so-called Umbral Dots (UDs)
and widely investigated earlier in a number of papers
\citep{2012ApJ...757..160J, 2016Ap&SS.361..366G,
2017AN....338..662E}. There are both as single UDs and UD chains.
When comparing the 1D oscillations power distribution
(Fig.~\ref{8}a) we found that the umbral vertical threadlike
periodicity sources to coincide with the location of the detected
UDs on the 2D map (Fig.~\ref{9}). The high-frequency oscillation
power here is higher than in the surrounding regions.

We superimposed the map for the $\sim$3-min periodicity at the
chromosphere level on the photospheric image, obtained at 18:36 UT
for area N1 (Fig.~\ref{9}). The comparison of the sources at various
levels of the sunspot atmosphere shows that the overlying
chromospheric sources (denoted by contours) are displaced relative
to the photospheric UDs. Their angular size is larger and has both a
compact and an extended shape. The photospheric chains of extended
UDs are accompanied by the same UDs chains at the chromosphere
level. Every source in the TiO corresponds to the source as a cell
or a filament in the H$\alpha$. This dependence indicates that the detected
oscillations are probably interconnected among themselves, and have
a common oscillation source located below, at the photosphere level.

To test the relationship between wave propagation for two UDs
sources located at different heights, we selected a bright
photospheric umbral dot shown by an arrow in Fig.~\ref{9}, and a
nearest oscillating chromospheric cell marked by cross in
Fig.~\ref{5}a. We filtered the brightness variations within the 2-4
min range to detect periodic oscillations and  to compare them. We
found that the level of oscillation power in the chromosphere is
much bigger than on the photosphere level. The oscillation peaks are
displaced relative to one another. The cross-correlation of the two
profiles showed an $\sim$20-sec lag with the coefficient $\sim$0.2.
The obtained lag value agrees with the earlier obtained values
\citep{2011SoPh..273...39K}  for the time difference of the wave
propagation at the photosphere-chromosphere levels. The weak level
of correlation, which was previously shown    in
\cite{2011A&A...525A..41K}, probably related to the displacement of
sources at different heights relative to one another. We found, near
the investigated cell, several pointed UDs that are, probably, the
footpoints of an inclined magnetic tube (or a tube bundle) that
expands with height. We observe the upper part of this magnetic
structure as an oscillating cell.

Based on the the obtained results we can conclude that for the
photosphere level, the sunspot oscillations have a broadband
behavior with a low level of high-frequency   oscillations in fine
umbral structures and strong low-frequency oscillations in the
penumbra. Both oscillations are related to the global p-mode
periodicity \citep{1982ApJ...253..386L, 1986ApJ...301..992L,
2006RSPTA.364..313B}. There is strong umbral oscillation damping
which was noted earlier in \cite{2009ApJ...706..909C} and
\cite{2011A&A...534A..65S} and interpreted as a local absorption
with emission decrease. Using helioseismology  methods
\citep{1990SoPh..129...83B, 2004SoPh..225..213N}, suppression of the
photospheric sunspot oscillations was recorded.

The periodic oscillations fill the whole umbra and are localized
with maximal power in cells and fine filaments. We revealed that all
oscillate in-phase, as a global standing wave without motions. In
\cite{2012ApJ...757..160J} a similar umbral oscillation behavior was
referred as a "drum skin" jitter. In the penumbra, the low-frequency
component prevails.

In the umbra there are brightenings which appear as photospheric
umbral dots. An increase in the periodic $\sim$3-min oscillations is
observed, as compared to the surrounding areas
\citep{2015ApJ...812L..15K, 2017ApJ...836...18C}. Those oscillations
are hidden mainly in smooth low-frequency brightness variations of
UDs.

We may assume that the observed increase of UD oscillations is a
result of the intensification of the background p-mode that
surrounds UDs. These oscillating areas can lead to the origin of
in-phase oscillations and wave motions in UDs.
\cite{2009A&A...501..735S} showed that, in the small-scale sources,
the magnetic field line curvature and the field strength can result
in a significant effect on the magnetoacoustic waves propagating.

There are spatial differences in localization of the 3-min sources
for TiO and H$\alpha$ spectral lines. When comparing  their position on the
1D coordinate-period diagrams (Fig.~\ref{5}a, Fig.~\ref{8}a) or a
narrowband map (Fig.~\ref{9}), we can see that at the photosphere
level the size of UDs is smaller then in the chromosphere. The
sources are displaced relative to each other. This agrees well with
\cite{2012ApJ...757..160J} where for the 4170 \AA ~continuum, H$\alpha$ and
Ca II 8542 \AA core was found from the spatial displacement of the
UDs. An explanation of such a behavior probably is the geometry of
magnetic field lines, where wave propagation is observed. There is a
magnetic field inclination relative to the solar normal at the
photosphere-chromosphere levels which increase the UD size due to
the expansion of magnetic tube with height.

\section{Conclusion}

We study the fine structure of oscillation sources at different
levels of the sunspot atmosphere. For the H$\alpha$ 
line (chromosphere level), the $\sim$3-min oscillations represent a
set of numerous independent various-amplitude oscillation sources as
a cells and filaments. The detected small angular size sources
weakly interact among themselves. There are only local oscillations.
Each spectrum harmonic corresponds to its narrowband source, and linked to the footpoints 
of elongated filamentous structures as waveguides, visible on the variance map.
Their shape varies from pointed in the umbral center to extended in the penumbra. 
These changes are related to the oscillation period increase. We show the presence of unstable
low-frequency $\sim$5 min oscillation sources in the umbra located
between high-frequency $\sim$3-min sources without overlap. Their
shape is mainly cellular. The temporal dynamics of the $\sim$3-min
oscillations show a non-monotonous character as low-frequency
trains. During the train evolution, the period drift is observed.
Those drifts were described earlier in the UV range
\citep{2012A&A...539A..23S}. The drifts are shown only for extended
sources with big angular size. A possible explanation for this is
instability in the oscillation period of small angular size regions
composing larger sources.

For the 7057 \AA (TiO) spectral  line (photosphere level), the
oscillations have a broadband character, with simultaneous, in-phase
temporal variations of the whole umbral region with $\sim$5-min
periodicity. We also found the existence of $\sim$3-min oscillations
in umbral dots with maximal power. We have shown that the spatial
positions of the $\sim$3-min oscillation sources have displacement
at different heights. A possible explanation for this is inclination
of the magnetic field lines, along which the waves propagate. The
increase in the angular size of the sources at the chromosphere
level is associated with the expansion  of the magnetic waveguide
with height.

The obtained results show that the umbral oscillation fine structures as cells and filaments 
are related to the footpoints of fine magnetic tubes, anchored in the umbra, where the waves propagation is observe.
These processes are reflection of slow magnetoacoustic wave propagation from the subphotospheric level to the corona. 
The sources localization and wave dynamics indicates a common source as broad-band subphotospheric
oscillations. It can be assumed that sunspot umbral and penumbral oscillations have different origins. 
In the umbra this is related to exciting oscillations in the sub-photospheric slow-wave resonator. 
In the penumbra, the origin of the filament components and their angular size increase are related 
to increase of the magnetic field inclination to the solar normal and to changes in the cutoff
frequency of the propagating waves.

\acknowledgments
This study was supported by the Ministry of Education and Science of the Russian Federation. The authors are grateful to the BBSO team for operating the instrument and performing the basic data reduction, and, especially, for the open data policy. We thank Dr. N.I. Kobanov for fruitful discussion, and Dr. S.A. Anfinogentov for his assistance in processing the experimental data. The study was performed within the basic funding from FR program II.16, RAS program KP19-270, and partially supported by the Russian Foundation for Basic Research (RFBR) under Grant 17-52-80064 BRICS-a. The BBSO operation is supported by NJIT and US NSF AGS-1821294 grant. GST operation is partly supported by the Korea Astronomy and Space Science Institute and Seoul National University and by the strategic priority research program of CAS with Grant No. XDB09000000.

\bibliographystyle{aasjournal} 
\bibliography{BibTeX_Sych} 

\end{document}